\documentclass[12pt]{article}
\usepackage{graphicx, amsmath, amssymb, cite, setspace, color,fancyhdr}
 \usepackage[top=1 in, bottom=1 in, left=1 in, right=1 in]{geometry}

\newcommand{\captionfonts}{\footnotesize} 
\makeatletter 
\long\def\@makecaption#1#2{%
  \vskip\abovecaptionskip
  \sbox\@tempboxa{{\captionfonts #1: #2}}%
  \ifdim \wd\@tempboxa >\hsize
    {\captionfonts #1: #2\par}
  \else
    \hbox to\hsize{\hfil\box\@tempboxa\hfil}%
  \fi
  \vskip\belowcaptionskip}
\makeatother

\def\fnote#1#2{\begingroup\def\thefootnote{#1}\footnote{#2}
     \addtocounter{footnote}{-1}\endgroup}

\begin{document}
\title{Populating the Whole Landscape}

\author{Adam~R.~Brown$^{1,2}$ \,and Alex~Dahlen$^{1}$ \vspace{.1 in}\\  
\vspace{-.3 em}  $^1$ \textit{\small{Physics Department, Princeton University, Princeton, NJ 08544, USA}}\\
\vspace{-.3 em}  $^2$ \textit{\small{Princeton Center for Theoretical Science, Princeton, NJ 08544, USA}}}
\date{}
\maketitle
\fnote{}{emails: \tt{adambro@princeton.edu, adahlen@princeton.edu}}

\begin{abstract}
\noindent Every de Sitter vacuum can transition to every other de Sitter vacuum despite any obstacle, despite intervening anti-de Sitter sinks, despite not being connected by an instanton. Eternal inflation populates the whole landscape.
\end{abstract}
\vspace*{12cm}
\maketitle
\pagebreak

We have no satisfactory theory of the initial conditions of the Universe. One of the great promises of eternal inflation is that it might make such a theory superfluous, because the predictions might be independent of initial conditions---that starting in any eternally inflating state the Universe would be driven towards the same attractor. 
In the context of the vast landscape of minima predicted by string theory, this would mean that the combination of exponential expansion of the vacua and  quantum transitions between the vacua would erase the memory of the starting point. 
The landscape would have no hair.

In order for this to be true,
a necessary condition is that the landscape be traversable, that there be a network of transitions connecting every inflating vacuum. The landscape must not break up into non-communicating superselection sectors, between which transitions are impossible, otherwise it would be necessary to know in which sector the Universe began. Any inflating vacuum must be able to populate the whole landscape.
There are two challenges to this idea.

\begin{figure}[h]
  \begin{center}
   \includegraphics[width=3.5in]{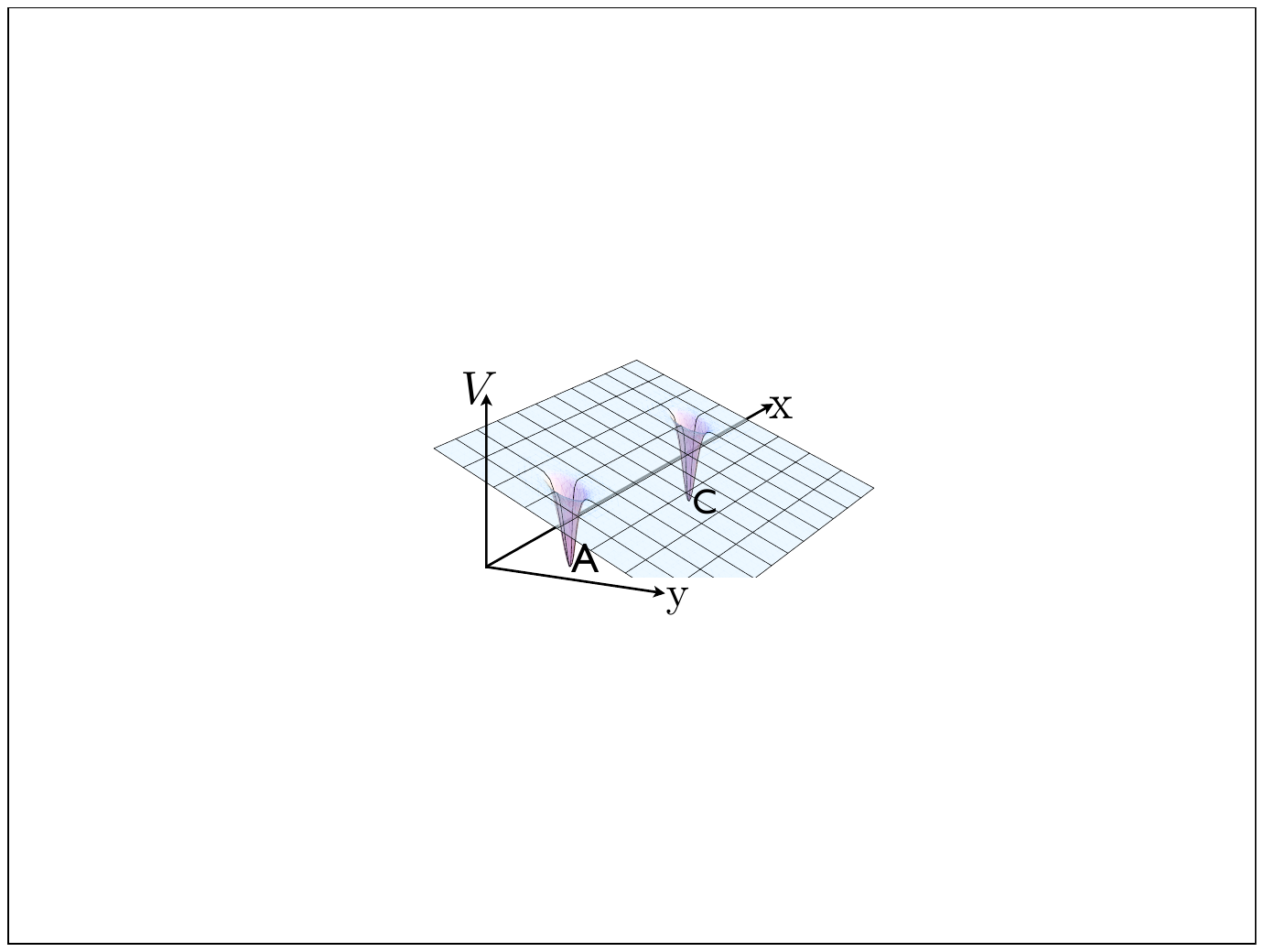} 
  \end{center}
  \caption{The potential, $V(x,y)$, as a function of the two fields, $x$ and $y$, has two minima on the line $y=0$ and a runaway direction as $y\rightarrow\infty.$ Coleman \cite{Coleman:1977py} showed that, if an instanton exists, its profile is given by the trajectory of a particle sliding in the inverse potential, $-V$, that starts at rest near $C$ and ends at rest at $A$. However, for the potential above, the minima are too far apart and the runaway direction is too steep, so there is no such trajectory---all the classical trajectories leaving $C$ end up sliding down the runaway direction.  There is no instanton that connects $A$ and $C$.}
     \label{fig:tilted}
\end{figure}

The first challenge is that not all vacua are connected by tunneling instantons. Instantons are Euclidean solutions that are used to calculate semiclassical tunneling rates \cite{Coleman:1977py}, and if there's no instanton, then it's not clear how to calculate the rate, and many authors have presumed that the rate is zero. Figure~\ref{fig:tilted} shows a simple example of minima not connected by an instanton, and similar examples are common in realistic landscapes  \cite{Cvetic:1994ya,Johnson:2008vn,Aguirre:2009tp,Brown:2011um}. In \cite{Johnson:2008vn}, it was argued that the absence of an instanton poses an obstacle to populating the whole landscape. 
 If the field starts in $A$, can it make it to $C$ even though there's no instanton? 

\begin{figure}[htbp] 
   \centering
   \includegraphics[width=3.5in]{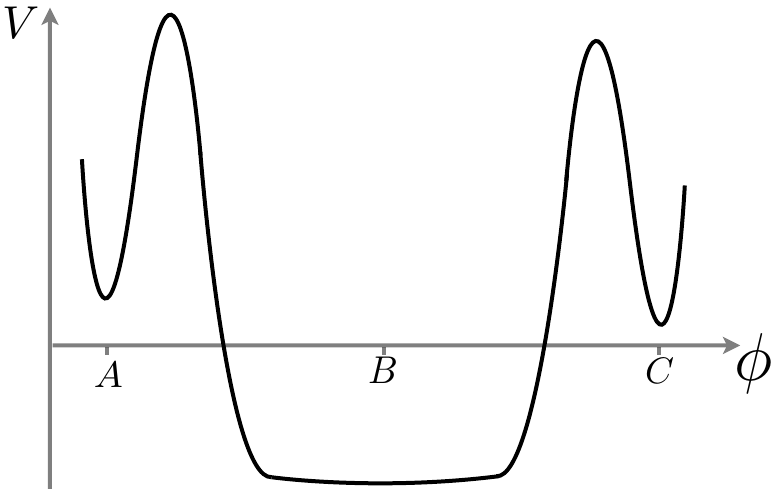}
 \caption{Again, Coleman's inverse-potential argument shows that there is no instanton that connects $A$ and $C$. Worse, this time every path from $A$ to $C$ must pass through a region of negative potential.  If the field settles in to $B$, it is stuck; $B$ is AdS and uptunneling is impossible.}
   \label{fig:dSadSdS}
\end{figure}

The second challenge is intervening
minima with negative potential. These minima give rise to anti-de Sitter (AdS) space, and act as sinks in the landscape. Once a field decays into AdS it can never leave---it can't uptunnel out, and it experiences a crunch in finite time. 
In \cite{Clifton:2007en} it was argued that intervening AdS sinks can divide the landscape into disconnected islands, and that only one island would get populated by eternal inflation.  A simple one-field example is shown in Fig.~\ref{fig:dSadSdS}.   If the field starts in $A$, can it make it to $C$ even though there's an AdS minimum in between?

In this paper, we argue that the answer to both  these questions is `yes', that every de Sitter vacuum can transition to every other de Sitter vacuum, that the \emph{whole} landscape gets populated. 
This is not a statement about the microphysical properties of the string-theory landscape, but rather a statement about tunneling---we argue that any (topologically connected) multi-minimum field-theory landscape is traversable. We show that the transition rate between any two de Sitter vacua, even without an instanton, is nonzero; a nonzero rate, no matter how tiny, is sufficient to drive the universe towards a single attractor.  In the final section, we provide evidence for a conjecture that even seemingly topologically disconnected landscapes are traversable in quantum gravity.
Let's start with non-relativistic quantum mechanics.

\subsection*{All transitions are possible in quantum mechanics}

 Consider  a particle with coordinates $(x,y)$ in the potential $V(x,y)$ from Fig.~1.  As in its quantum field theory counterpart, there is no tunneling instanton connecting the $A$ and $C$ vacua.  Nevertheless if the particle starts in $A$, there is a non-zero amplitude to reach $C$. The Schr\"{o}dinger equation is a diffusion equation with the property that after an infinitesimal time the wavefunction has support everywhere that is not infinitely removed from the starting point (except at a measure zero set of nodes).

Here's another way to say it: In the path integral approach, the wavefunction evolves along all paths with an amplitude weighted by $\exp{[i S]}$.  In the semiclassical approximation, the path integral is 
dominated by classical paths; far from the classical paths, 
 it is argued that the integrand is oscillating so rapidly that the integral cancels.  Beyond the semiclassical approximation, however, this cancellation is not perfect; except for a measure zero set of nodes that come from complete destructive interference, the wavefunction will be non-zero everywhere that is accessible by a Lorentzian path with finite $S$.  Only when $S\rightarrow\infty$ is the destructive interference perfect and uniform (by the Riemann-Lebesgue lemma); a quantum mechanical particle can tunnel anywhere, except through an infinite barrier.

Our diagnostic in determining what counts as a transition is that, first, there is a non-zero probability to be measured in $C$ and that, second, once measured in $C$, the particle classically endures there (though semiclassically it may be unstable to further decay). This will be the appropriate diagnostic when it comes to the full landscape, because the gravitational degrees of freedom will decohere the different branches of the wavefunction with different values of $\phi$ and in so doing play the role of the measurement apparatus. 
 
\subsection*{Some transitions are impossible in flat spacetime}

\begin{figure}[htbp] 
   \centering
   \includegraphics[width=2.8in]{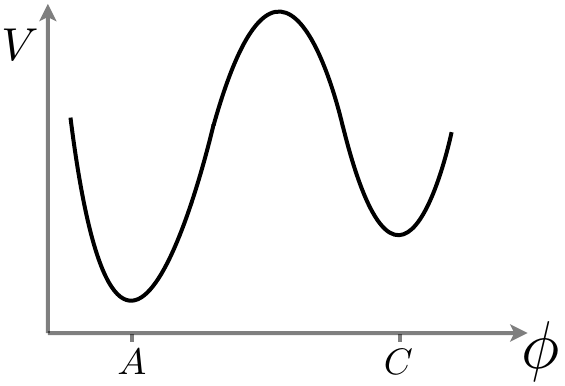}
   \caption{Uptunneling is forbidden for quantum field theory in Minkowski space, but permitted for quantum mechanics, quantum field theory in a finite box, and quantum field theory in de Sitter space.}
   \label{fig:uptunneling}
\end{figure}

For quantum field theory in a fixed Minkowski space (with $G_\text{N}=0$), there are some truly impossible transitions.  One simple example is uptunneling; a field localized in the $A$ minimum  of Fig.~\ref{fig:uptunneling} will never be able to uptunnel to the $C$ minimum. Though we might measure a finite region of space to be momentarily in $C$, this does not count as a transition because the region of $C$ does not classically endure: it contracts inwards and, within a few light-crossing times, disappears entirely.  

Transitions are likewise impossible from $A$ to $C$ in the potential of Fig.~1.  Again, if a finite region of $C$ embedded in $A$ does form, it will not endure: the wall will thicken, and the field inside the wall will slide off in the runaway direction,  bringing the region of $C$ with it.  The issue in  our examples is not just that there is no instanton, it's that in Minkowski space there is no way to make a field configuration that both asymptotes to $A$ at spatial infinity and contains a classically enduring region of $C$. 

We have a small paradox here.
Quantum field theory is the non-relativistic quantum mechanics of a field (with a relativistic dispersion relation), and we just argued that in non-relativistic quantum mechanics, the wavefunction will immediately spread out over all configuration space.  What goes wrong with this argument?

The problem is that transitioning from $A$ to $C$ requires tunneling through an infinite energy barrier.  In quantum field theory, the wavefunction gives a probability distribution over field configurations and lives in a potential $U[\phi(\vec{x})]$ defined by
\begin{equation}
U[\phi(\vec{x})]=\int d^3 x  \left\{ \frac12 (\vec{\nabla} \phi(\vec{x}))^2 + V(\phi(\vec{x})) \right\} .
\label{Udef}
\end{equation}
The field should be thought of as tunneling not through $V$ but through $U$. 
In the examples above, the only way to form an enduring region of $C$ is to deform an infinite volume  of field away from its minimum, so there is an infinite  barrier $U[\phi(\vec{x})]$ to tunneling.  It is the infinite volume of flat space that prevents these transitions.

\subsection*{All transitions are possible with a finite volume}

Quantum fields in a finite periodic box, however, can get from $A$ to $C$. This follows directly from the fact that all transitions are possible in quantum mechanics and that the zero mode of the field can be treated as a quantum mechanical particle.  It is always allowed, therefore, that the field everywhere in the whole box transitions homogeneously through the barrier.   The rate for this process goes to zero as the box grows infinitely large, as can be seen using the WKB approximation.  Both the mass of the zero mode and the potential faced by the zero mode are proportional to the volume; the WKB integrand is $\sqrt{2m U}$, so the rate to homogeneously transition is given by $\exp( - 2 \, \textrm{volume} \int d \phi  \sqrt{2V})$.

A fixed de Sitter background (with $G_\text{N}=0$) is like a finite box in the sense that only a finite region of field needs to make the transition---inflation ensures that, once formed, a horizon-sized volume of $C$ will classically endure.  Indeed, field theory on the causal patch of de Sitter is just the field theory of a horizon-sized box (with a finite temperature and a position-dependent Hamiltonian \cite{Brown:2007sd}). Therefore, on a fixed de Sitter background, all transitions are possible.

The crucial feature of de Sitter is that only a finite volume needs to transition, not that there is a nonzero temperature.  Turning on a temperature cannot make an impossible transition possible.  For instance, a finite box at zero temperature can uptunnel, but hot Minkowski cannot; as we argued above, Minkowski faces an infinite potential barrier $U$ to uptunneling, which a finite $T$ cannot overcome.  (A temperature can, however, enhance the rate of an already-possible transition \cite{Linde:1981zj,Lee:1987qc}.)

\subsection*{The whole landscape gets populated}

Consider the landscape of Fig.~\ref{fig:dSadSdS} with gravity dynamical ($G_\text{N}\neq0$). This potential encompasses both of the challenges of the introduction, and we have constructed it to serve as a proxy for the most difficult transitions in the landscape. 

In a fixed de Sitter space, there would be an instanton that tunnels from $A$ to $B$, and then a second that tunnels from $B$ to $C$, so that two sequential semiclassical tunneling events would populate $C$. But gravitational backreaction blocks this route.  Were $V_B$ zero, a field that settles in to $B$ would find itself in Minkowski space, which cannot uptunnel. Since $V_B$ is negative, not only can the field not uptunnel, it cannot even `settle in' since it will crunch in finite time.

\begin{figure}[htbp] 
   \centering
   \includegraphics[width=4.8in]{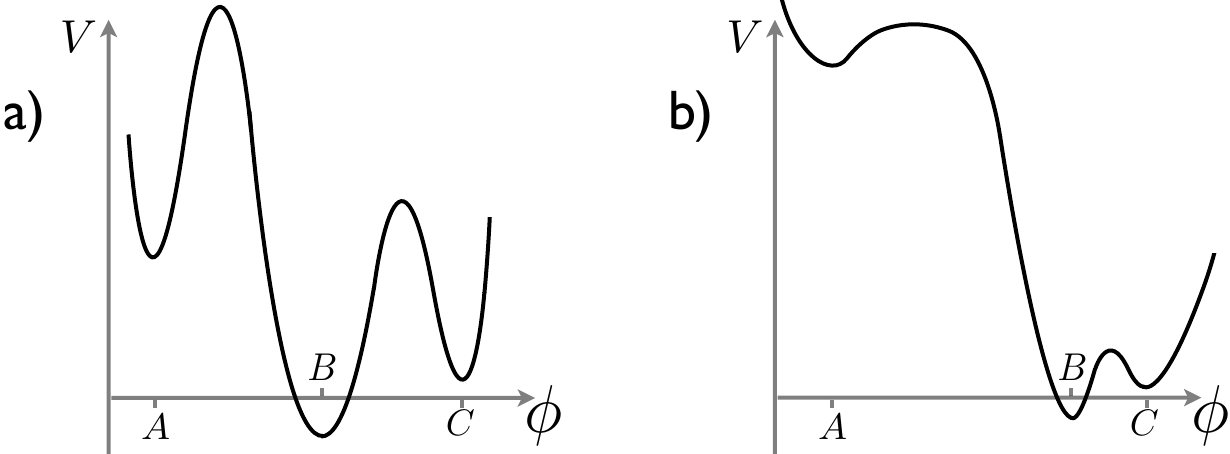}
   \caption{Two potentials for which $C$ is reachable using instantons.  a) This potential has the $C$ vacuum close enough in field space and has the right hierarchy between $V_A$, $V_B$ and $V_C$ to allow the collision of two bubbles of $B$ to produce an enduring region of $C$. b) This potential has a Hawking-Moss instanton that carries the field homogeneously to the top of the $AB$ barrier. As the field rolls off the crest it gathers so much velocity that it overshoots $B$ and ends up in $C$. The standard decay instanton from $A$ leads not to $B$ but to $C$.}
   \label{fig:sometimespotentials}
\end{figure}

One method of traversing an AdS minimum that works for some potentials is bubble collisions \cite{Easther:2009ft,Johnson:2010bn}.  In the potential of Fig.~\ref{fig:sometimespotentials}a, if the field begins uniformly in $A$, tunneling events will produce bubbles of $B$ that expand and eventually collide. These collisions can throw the field into the $C$ vacuum, which begins eternally inflating---one bubble nucleation can't reach $C$ but two bubbles colliding can.  This collision mechanism only works, however, when the potential obeys several constraints \cite{Johnson:2010bn}, amongst them that the $C$ vacuum has lower energy than the $A$ vacuum, and that the $C$ vacuum is close in field space.  We have constructed the potential of Fig.~\ref{fig:dSadSdS} so that the $C$ vacuum is too distant and cannot be populated by bubble collisions.

 Another method of traversing an AdS minimum that works for some potentials is a large field velocity.  The instanton that mediates tunneling out of $A$ deposits the field somewhere on the slope between $A$ and $B$.  (The extreme example of this is the Hawking-Moss instanton \cite{Hawking:1981fz}, for which the field emerges uniformly at the top of the $AB$ barrier.  This instanton describes a thermal process by which the field is given just enough of a homogeneous push to reach the crest of the barrier with zero velocity.)  After emerging, the field rolls down the slope towards $B$, picking up velocity.  In the potential of Fig.~\ref{fig:sometimespotentials}b, this velocity is large by the time the field reaches the bottom and, as we will soon show, a large enough field velocity can carry the field through $B$, over the $BC$ hump, and into $C$.  However, we have constructed the potential of Fig.~\ref{fig:dSadSdS} with a broad $B$ minimum and a large hump between $B$ and $C$ so that this mechanism cannot reach $C$.\footnote{A Hawking-Moss (HM) instanton that is perched on the crest between $B$ and $C$ does not represent decay from $A$ and so cannot mediate the desired transition.  In the language of Starobinsky \cite{Starobinsky}, a stochastic random walk of the field  would get stuck in $B$ \cite{Clifton:2007en}.  In the language of Jensen and Steinhardt \cite{Jensen:1988zx}, the HM instanton is the limit of Coleman-De Luccia (CDL) instantons as the wall thickness grows to be horizon size, but the CDLs of which the HM is the limit connect not $A$ and $C$ but $B$ and $C$ \cite{Weinberg:2006pc}.} 
 
There is no instanton or chain of instantons that connects $A$ and $C$. However, as we argued in the quantum mechanics section, this does not mean that the transition is impossible, just that it must proceed by non-instanton means. We will show that the rate is nonzero by finding a finite action Lorentzian path that connects the state where the field is uniformly in $A$ to a state with a region of $C$ large enough to classically endure.

In constructing this path, we are not required to choose a $\phi(t)$ that satisfies the scalar field equation, which for an FRW metric is
\begin{equation}
\ddot{\phi} + 3 H \dot{\phi}  =  - \frac{d V} {d \phi}, \label{eq:field}
\end{equation}
because the path contributes to the path integral regardless.   However, once $\phi(t)$ is chosen, we have no further freedom to choose the evolution of the scale factor $a(t)$, because it is set by the gravitational constraint equation.  Every path that contributes to the path integral must obey Friedmann's equation, which  on O(4)-symmetric slices is
\begin{equation}
H^2 \  \equiv  \   \frac{\dot{a}^2}{a^2} =   -\frac{1}{a^2} + \frac{1}{3M_\text{Pl}^2}\left( \frac{1}{2} \dot{\phi}^2 + V \right). \label{Fried}
\end{equation}
We must therefore take care to ensure that the path is well behaved as it crosses the region with $V<0$, because if $a$ diverges then so does $S$, and if spacetime crunches then the field never reaches $C$.  

There are in fact many satisfactory paths.  As an existence proof, we will describe one such path---a relative of the Hawking-Moss process---that has a clear physical interpretation because it gets all its non-classical behavior in at the start.
In the Hawking-Moss process, the field leaves the crest of the $AB$ barrier with zero velocity and, for the potential of Fig.~\ref{fig:dSadSdS}, does not gather enough velocity rolling down to make it past $B$. Consider the path that gives the field a bigger initial push; a large enough uniform positive $\dot\phi$ will propel the field straight past $B$, and on to $C$.

To confirm this story, we need to check that a large enough initial field velocity can propel the field arbitrarily far. 
In the limit of huge $\dot{\phi}$, Eqs.~\ref{eq:field} and \ref{Fried} become
\begin{equation}
\dot{\phi} = \sqrt{6} M_\text{Pl} H  =   \sqrt{\frac{2}{3}} M_\text{Pl} \frac{1}{t} ,
\end{equation}
so that so long as the desired $\Delta \phi$ is sub-Planckian, and the $V$s are sub-Planckian, the field can always get there with a $\dot{\phi}_{\textrm{initial}}$ that is itself sub-Planckian:
\begin{equation}
\Delta \phi  =  \sqrt{\frac{2}{3}} M_\text{Pl} \Delta \log t  = -  \sqrt{\frac{2}{3}} M_\text{Pl} \Delta \log \dot{\phi}.
\end{equation}
The scale factor is well behaved throughout and indeed the spacetime expands by less than an e-fold: 
\begin{equation}
N_\textrm{e-folds} = \int da/a = \int H dt  = \Delta\phi/(\sqrt6 M_\text{Pl}).
\end{equation}
(For some potentials, the large $\dot\phi$ means the field classically overshoots not just $B$ but also the target minimum $C$, eventually coming to rest elsewhere in the landscape.  For these potentials, we need two quantum violations of the equation of motion, Eq.~\ref{eq:field}: one to propel the field as far as $C$, and another to stop it on upon arrival. For other potentials with super-Planckian field excursions $\Delta \phi$, we require repeated quantum violations along the path, many small pushes, to ensure the energy densities remain always sub-Planckian. Such paths still obey the constraint equation, Eq.~\ref{Fried}, so they still contribute to the path integral and $A$ can still transition to $C$.)

The original interpretation of the Hawking-Moss process was that it describes the transitioning of the entire Universe to the top of the barrier---the whole equator of de Sitter fluctuates up and none of the original vacuum remains \cite{Hawking:1981fz, Coleman:1980aw}. Following this interpretation, the transition path described above corresponds to a complete closed slice of de Sitter uniformly transitioning from $A$ to $C$. However, a  modern interpretation of Hawking Moss \cite{Brown:2007sd} (and the one we prefer) is that the process describes only a single horizon volume of field perched uniformly atop the barrier, and that outside the horizon the field smoothly interpolates back towards the false vacuum.  Following this  modern interpretation, the transition path described above corresponds to the formation of a horizon-sized region of $C$, outside of which the field smoothly interpolates through $B$ and back towards $A$. Though the region of $B$ does grow in both directions, consuming $A$ and $C$ alike (and eventually crunching), the region of $C$ is large enough, and inflating fast enough, that it classically endures.\footnote{Under either interpretation, both the Hawking-Moss process and the tunneling path described above consume more than a horizon volume of $A$. As a result, neither process is possible from Minkowski space (or from AdS space), because of the infinite horizon size. Farhi, Guth and Guven \cite{Farhi:1989yr} have proposed an alternative to the Hawking-Moss/Coleman-de Luccia process that consumes less than a horizon volume of $A$. In this paper we have assumed, following \cite{Freivogel:2005qh}, that this instanton does not contribute to the path integral. If in fact it does, then presumably there is an analogous alternative to our process that also consumes less than a horizon volume of $A$. If valid, the FGG mechanism would make populating the landscape unchallenging.}

In a \emph{fixed} de Sitter space the frequency of such fluctuations is given by  the Boltzmann suppression factor
\begin{equation}
\Gamma \sim  \exp \left[  - \frac{\Delta U}{T_{\textrm{dS}} }  \right]  \sim  \exp \left[  - \frac{8 \pi^2}{3 H^4} \left( \frac{1}{2} \dot{\phi}^2 \right)    \right] . 
\label{bunchdavies} 
\end{equation}
This rate will generally be an underestimate, first because it does not include gravitational backreaction, which will tend to shrink the horizon volume and increase the effective temperature, and second because, depending on the details of the potential, this may not be the most efficient route across the barrier. The important thing, however, is that the rate is non-zero: a non-zero rate, no matter how small, is sufficient to erase memory of initial conditions and populate the whole landscape.

\subsection*{Discussion}

We have described one finite-action Lorentzian path from $A$ to $C$, but there are many. A final concern is that their amplitudes might add  to zero, they might completely destructively interfere.  This will indeed happen for some final states, but the set of such final states will be measure zero.   The wavefunction provides a map from configuration space into $\mathbb{C}$; because the origin is only a single point in the complex plane, nodes of the wavefunction generically form a codimension two surface.  Even if the state with the field uniformly in $C$ is inaccessible, there will be an accessible state arbitrarily close in configuration space.

So far, our arguments have been about topologically connected field-theory landscapes, where any field value can be smoothly deformed into any other.  However, it seems reasonable to conjecture that in quantum gravity all vacua, even the seemingly disconnected ones, are reachable from a de Sitter starting point.  There are three pieces of supporting evidence. 
First, in string theory, a persistent pattern of discovery is that seemingly disconnected landscape areas are in fact connected \cite{Greene:1996cy, Chialva:2007sv}.
Second, even in the context of Coleman-De Luccia instantons, de Sitter vacua can transition across  \emph{infinite} potential barriers---as $V \rightarrow \infty$ the instanton that mediates the transition shrinks to zero size, so that the instanton action goes to zero and decay proceeds with a nonzero rate of once per Poincar\'e recurrence time. So, insofar as disconnected areas of field space can be thought of as separated by an infinite barrier, they will still be reached. 
Finally, there is good evidence for an equivalent conjecture as it relates to global charges, that non-perturbative gravitational effects violate all global symmetries \cite{Kallosh:1995hi,Kamionkowski:1992mf,Abbott1989687} and conserve no global charges, so that quantum gravity respects no global superselection sectors.  

In this paper we have shown that, as long as there is a nonzero amplitude to eternally inflate, the whole landscape gets populated. This amplitude will be nonzero if the initial conditions of the Universe are de Sitter, but it will also be nonzero if the initial conditions are defined on any compact surface. Finite spaces can transition everywhere,  so if the Universe starts finite, some branch of the wavefunction will uptunnel and inflate. The only way to stop eternal inflation is to start with a spatially infinite Universe.  If any part of the landscape supports eternal inflation, then the Universe is necessarily infinite in space or time. 

\subsection*{Acknowledgements}
Thanks to Nima Arkani-Hamed, Ben Freivogel, Alan Guth, Kurt Hinterbichler, Matt Johnson, Navin Sivanandam, Paul Steinhardt, Lenny Susskind, and Erick Weinberg. 

\pagebreak

\bibliographystyle{utphys}
\bibliography{mybib.bib}

\providecommand{\href}[2]{#2}\begingroup\raggedright\begin{thebibliography}{10}

\bibitem{Coleman:1977py}
S.~R. Coleman, ``{The Fate of the False Vacuum. 1. Semiclassical Theory},''
\href{http://dx.doi.org/10.1103/PhysRevD.15.2929}{{\em Phys. Rev.} {\bfseries
  D15} (1977) 2929--2936}.

\bibitem{Cvetic:1994ya}
M.~Cvetic and H.~H. Soleng, ``{Naked Singularities in Dilatonic Domain Wall
  Spacetimes},'' \href{http://dx.doi.org/10.1103/PhysRevD.51.5768}{{\em Phys.
  Rev.} {\bfseries D51} (1995) 5768--5784},
\href{http://arxiv.org/abs/hep-th/9411170}{{\ttfamily arXiv:hep-th/9411170}}.

\bibitem{Johnson:2008vn}
M.~C. Johnson and M.~Larfors, ``{An Obstacle to Populating the String Theory
  Landscape},'' \href{http://dx.doi.org/10.1103/PhysRevD.78.123513}{{\em Phys.
  Rev.} {\bfseries D78} (2008) 123513},
\href{http://arxiv.org/abs/0809.2604}{{\ttfamily arXiv:0809.2604 [hep-th]}}.

\bibitem{Aguirre:2009tp}
A.~Aguirre, M.~C. Johnson, and M.~Larfors, ``{Runaway Dilatonic Domain
  Walls},'' \href{http://dx.doi.org/10.1103/PhysRevD.81.043527}{{\em Phys.
  Rev.} {\bfseries D81} (2010) 043527},
\href{http://arxiv.org/abs/0911.4342}{{\ttfamily arXiv:0911.4342 [hep-th]}}.

\bibitem{Brown:2011um}
A.~R. Brown and A.~Dahlen, ``{The Case of the Disappearing Instanton},''
\href{http://arxiv.org/abs/1106.0527}{{\ttfamily arXiv:1106.0527 [hep-th]}}.

\bibitem{Clifton:2007en}
T.~Clifton, A.~D. Linde, and N.~Sivanandam, ``{Islands in the Landscape},''
  \href{http://dx.doi.org/10.1088/1126-6708/2007/02/024}{{\em JHEP} {\bfseries
  02} (2007) 024},
\href{http://arxiv.org/abs/hep-th/0701083}{{\ttfamily arXiv:hep-th/0701083}}.

\bibitem{Brown:2007sd}
A.~R. Brown and E.~J. Weinberg, ``{Thermal Derivation of the Coleman-De Luccia
  Tunneling prescription},''
  \href{http://dx.doi.org/10.1103/PhysRevD.76.064003}{{\em Phys. Rev.}
  {\bfseries D76} (2007) 064003},
\href{http://arxiv.org/abs/0706.1573}{{\ttfamily arXiv:0706.1573 [hep-th]}}.

\bibitem{Linde:1981zj}
A.~D. Linde, ``{Decay of the False Vacuum at Finite Temperature},''
\href{http://dx.doi.org/10.1016/0550-3213(83)90293-6}{{\em Nucl. Phys.}
  {\bfseries B216} (1983) 421}.

\bibitem{Lee:1987qc}
K.-M. Lee and E.~J. Weinberg, ``{Decay of the True Vacuum in Curved
  Space-time},''
\href{http://dx.doi.org/10.1103/PhysRevD.36.1088}{{\em Phys. Rev.} {\bfseries
  D36} (1987) 1088}.

\bibitem{Easther:2009ft}
R.~Easther, J.~T. Giblin, Jr, L.~Hui, and E.~A. Lim, ``{A New Mechanism for
  Bubble Nucleation: Classical Transitions},''
  \href{http://dx.doi.org/10.1103/PhysRevD.80.123519}{{\em Phys. Rev.}
  {\bfseries D80} (2009) 123519},
\href{http://arxiv.org/abs/0907.3234}{{\ttfamily arXiv:0907.3234 [hep-th]}}.

\bibitem{Johnson:2010bn}
M.~C. Johnson and I.-S. Yang, ``{Escaping the Crunch: Gravitational Effects in
  Classical Transitions},''
  \href{http://dx.doi.org/10.1103/PhysRevD.82.065023}{{\em Phys. Rev.}
  {\bfseries D82} (2010) 065023},
\href{http://arxiv.org/abs/1005.3506}{{\ttfamily arXiv:1005.3506 [hep-th]}}.

\bibitem{Hawking:1981fz}
S.~W. Hawking and I.~G. Moss, ``{Supercooled Phase Transitions in the Very
  Early Universe},''
\href{http://dx.doi.org/10.1016/0370-2693(82)90946-7}{{\em Phys. Lett.}
  {\bfseries B110} (1982) 35}.

\bibitem{Starobinsky}
A.~A. Starobinsky in {\em {Current Topics in Field Theory, Quantum Gravity and
  Strings}}.
\newblock Springer, Heidelberg, 1986.

\bibitem{Jensen:1988zx}
L.~G. Jensen and P.~J. Steinhardt, ``{Bubble Nucleation for Flat Potential
  Barriers},''
\href{http://dx.doi.org/10.1016/0550-3213(89)90539-7}{{\em Nucl. Phys.}
  {\bfseries B317} (1989) 693--705}.

\bibitem{Weinberg:2006pc}
E.~J. Weinberg, ``{Hawking-Moss Bounces and Vacuum Decay Rates},''
  \href{http://dx.doi.org/10.1103/PhysRevLett.98.251303}{{\em Phys. Rev. Lett.}
  {\bfseries 98} (2007) 251303},
\href{http://arxiv.org/abs/hep-th/0612146}{{\ttfamily arXiv:hep-th/0612146}}.

\bibitem{Coleman:1980aw}
S.~R. Coleman and F.~De~Luccia, ``{Gravitational Effects on and of Vacuum
  Decay},''
\href{http://dx.doi.org/10.1103/PhysRevD.21.3305}{{\em Phys. Rev.} {\bfseries
  D21} (1980) 3305}.

\bibitem{Farhi:1989yr}
E.~Farhi, A.~H. Guth, and J.~Guven, ``{Is it Possible to Create a Universe in
  the Laboratory by Quantum Tunneling?},''
\href{http://dx.doi.org/10.1016/0550-3213(90)90357-J}{{\em Nucl. Phys.}
  {\bfseries B339} (1990) 417--490}.

\bibitem{Freivogel:2005qh}
B.~Freivogel, V.~E. Hubeny, A.~Maloney, R.~C. Myers, M.~Rangamani, and
  S.~Shenker, ``{Inflation in AdS/CFT},''
  \href{http://dx.doi.org/10.1088/1126-6708/2006/03/007}{{\em JHEP} {\bfseries
  03} (2006) 007},
\href{http://arxiv.org/abs/hep-th/0510046}{{\ttfamily arXiv:hep-th/0510046}}.

\bibitem{Greene:1996cy}
B.~R. Greene, ``{String Theory on Calabi-Yau Manifolds},''
\href{http://arxiv.org/abs/hep-th/9702155}{{\ttfamily arXiv:hep-th/9702155}}.

\bibitem{Chialva:2007sv}
D.~Chialva, U.~H. Danielsson, N.~Johansson, M.~Larfors, and M.~Vonk,
  ``{Deforming, Revolving and Resolving---New Paths in the String Theory
  Landscape},'' \href{http://dx.doi.org/10.1088/1126-6708/2008/02/016}{{\em
  JHEP} {\bfseries 02} (2008) 016},
\href{http://arxiv.org/abs/0710.0620}{{\ttfamily arXiv:0710.0620 [hep-th]}}.

\bibitem{Kallosh:1995hi}
R.~Kallosh, A.~D. Linde, D.~A. Linde, and L.~Susskind, ``{Gravity and Global
  Symmetries},'' \href{http://dx.doi.org/10.1103/PhysRevD.52.912}{{\em Phys.
  Rev.} {\bfseries D52} (1995) 912--935},
\href{http://arxiv.org/abs/hep-th/9502069}{{\ttfamily arXiv:hep-th/9502069}}.

\bibitem{Kamionkowski:1992mf}
M.~Kamionkowski and J.~March-Russell, ``{Planck Scale Physics and the
  Peccei-Quinn Mechanism},''
  \href{http://dx.doi.org/10.1016/0370-2693(92)90492-M}{{\em Phys. Lett.}
  {\bfseries B282} (1992) 137--141},
\href{http://arxiv.org/abs/hep-th/9202003}{{\ttfamily arXiv:hep-th/9202003}}.

\bibitem{Abbott1989687}
L.~F. Abbott and M.~B. Wise, ``Wormholes and global symmetries,''
  \href{http://dx.doi.org/DOI: 10.1016/0550-3213(89)90503-8}{{\em Nuclear
  Physics B} {\bfseries 325} no.~3, (1989) 687 -- 704}.

\end{thebibliography}\endgroup

\end{document}